\documentclass[symmetry,article,submit,oneauthor,pdftex]{mdpi} 
\pdfoutput=1
\newcommand{\jpet}{\mbox{J-PET}}

\usepackage[super]{nth}

\firstpage{1} 
\pubvolume{xx}
\issuenum{1}
\articlenumber{5}
\pubyear{2020}
\copyrightyear{2020}
\history{Received: date; Accepted: date; Published: date}

\Title{Sensitivity of discrete symmetry tests in the positronium system with the J-PET detector}

\Author{
  Aleksander Gajos $^{1}$*
}

\address[1]{%
  $^{1}$ \quad Faculty of Physics, Astronomy and Applied Computer Science, Jagiellonian University,  S.~Łojasiewicza 11, 30-348 Kraków, Poland
}

\corres{Correspondence: aleksander.gajos@uj.edu.pl}

\abstract{%
  Study of certain angular correlations in the three-photon annihilations
  of the triplet state of positronium, the electron-positron bound state,
  may be used as a probe of potential CP and CPT-violating effects 
  in the leptonic sector. 
  We present the perspectives of CP and CPT tests using this process recorded with
  a novel detection system for photons in the positron annihilation energy range,
  the Jagiellonian PET~(\jpet{}).
  We demonstrate the capability of this system to register three-photon annihilations
  with an unprecedented range of kinematical configurations and 
  to measure the CPT-odd  correlation between positronium spin and annihilation plane orientation
  with a precision improved by at least an order of magnitude with respect to present results.
  We also discuss the the means to control and reduce detector asymmetries in order to allow
  \jpet{} to set the first measurement of the correlation between positronium spin and
  momentum of the most energetic annihilation photon which has never been studied to date.
}

\keyword{discrete symmetry, CPT, positronium}

\begin{document}

\section{Introduction}

The notion of searching of for violation of fundamental discrete symmetries in purely leptonic systems and in electromagnetic interactions, although not new,
has been outside of the mainstream of symmetry tests in physics  
for few decades after first violation had been discovered.

While the pioneering discoveries naturally led the attention towards weak interactions~\cite{PhysRev.105.1413, PhysRevLett.13.138},
other violation mechanisms should not be ruled out precipitately.

A viable purely leptonic system for tests of discrete symmetries is constituted by positronium exotic atoms.
As a bound state of electron and positron, positronium is the lightest matter-antimatter system
and at the same time an eigenstate of the C and P operations,
making it an ideal candidate for searching of symmetry violating efects~\cite{moskal_potential}.
This potential has been recognized already in 1967 by Mills and Berko,
who performed a search for decays of annihilations of the positronium C-even singlet state
into a C-odd three-photon final state, concluded with a null result~\cite{PhysRevLett.18.420}.

The field of discrete symmetry studies in the lepton sector has seen little activity until Bernreuther \textit{et al.} pointed out that violations of the CP and CPT
could be manifested by certain non-vanishing angular correlations in the decays of positronium atoms
if the corresponding operators constructed with observables available in the decay
are odd under a given symmetry transformation~\cite{Bernreuther:1988tt}.
Several implementations of tests based on such angular correlations followed,
with the best measurements to date yielding results consistent with conservation of both
CP and CPT with precision at the level of $10^{-3}$~\cite{cp_positronium, cpt_positronium}.
Notably, authors of the most recent CPT test performed using the Gammasphere array of germanium detectors have also extended the searches for C violation by searching for higher-order C-prohibited annihilations of positronium~\cite{PhysRevA.66.052505}. Other prohibited positronium decays were studied to test lowest-order QED calculations~\cite{PhysRevA.54.1947, VONBUSCH1994300}.

Since the discovery of neutrino oscillations, searches for leptonic CP violation
were strongly concentrated on neutrino physics
~\cite{RevModPhys.84.515}.
Although other tests were attempted e.g.\ by searching for the electric dipole moment of $\tau$~\cite{INAMI200316}, it is indeed the long-baseline neutrino oscillation experiments which first show hints of observation of CP violation at $3\sigma$~level~\cite{PhysRevLett.121.171802, PhysRevD.98.032012, Abe:2019vii}.

The interest in positronium as a potential probe of CPT violation has been recently revived
by the postulation of possible effects of Lorentz invariance observable with positronium
in the framework of the Standard Model Extension (SME).
SME is a general-realistic effective field theory of Lorentz violation,
which extends systems’ Lagrangians to include all effectively possible Lorentz-violating terms.
The inherent relation of Lorentz and CPT invariance allows for defining searches of violations of the latter in terms of SME paramters' measurement.
A number of possible experiments based on hyperfine spectroscopy of positronium have been postulated using both minimal SME~\cite{Kostelecky:2015nma} and non-minimal SME scenarios~\cite{Vargas:2019swg}.

The \jpet{} collaboration strives to explore an experimental programme complementary to
the SME-motivated spectroscopic studies.
Exploiting the potential of a novel powerful detector of photons
in the positron annihilation energy range, we aim at extending the measurements of angular correlations
in the decays of the positronium triplet state,
which are sensitive to effects violation of fundamental symmetries~\cite{moskal_potential}.
In this work, we present the scope of experimental CP and CPT tests available with the \jpet{} detector
based on large-acceptance exclusive detection of ortho-positronium annihilations and
an unconventional scheme of positronium spin orientation estimation on a single-event basis.

\section{The \jpet{} detector}\label{sec:detector}
\jpet{} was conceived as the first Positron Emission Tomography (PET) scanner
based on plastic scintillators~\cite{Moskal:2014sra, Moskal:2014rja, Niedzwiecki:2017nka, Kowalski:2018jra}.
While actively exploited in medical imaging research towards constructing
a cost-effective whole-body PET scanner~\cite{Niedzwiecki:2017nka, 8824622, pet_clinics}
and devising new imaging modalities such as spatially-resolved determination 
of properties positronium atoms produced during a PET scan~\cite{Moskal:2018wfc, Moskal:2018wfc, Moskal:Nature:2019, Moskal:2019nqk},
\jpet{} also constitutes a robust detector of photons in the sub-MeV range,
well suited for studies of phenomena such as positronium annihilation
and entanglement of photons in the field of fundamental research~\cite{Hiesmayr:2017xgx, Moskal:2018pus, Hiesmayr:2018rcm}.

\begin{figure}[hbt]
  \centering
  \includegraphics[width=8cm]{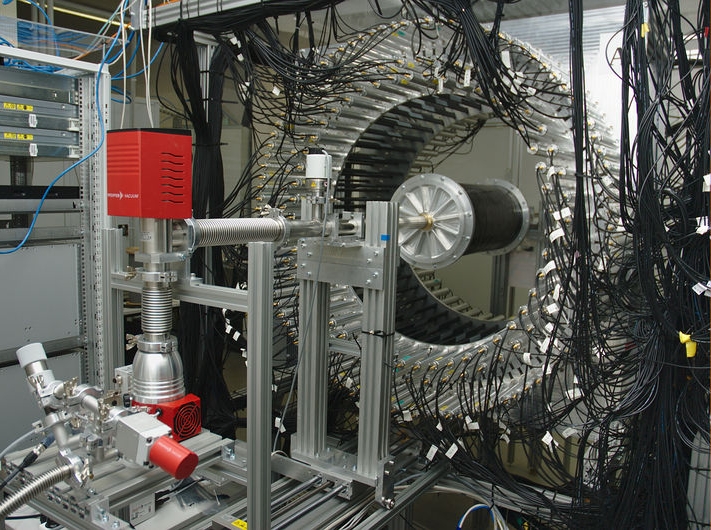}
    \includegraphics[width=6.5cm]{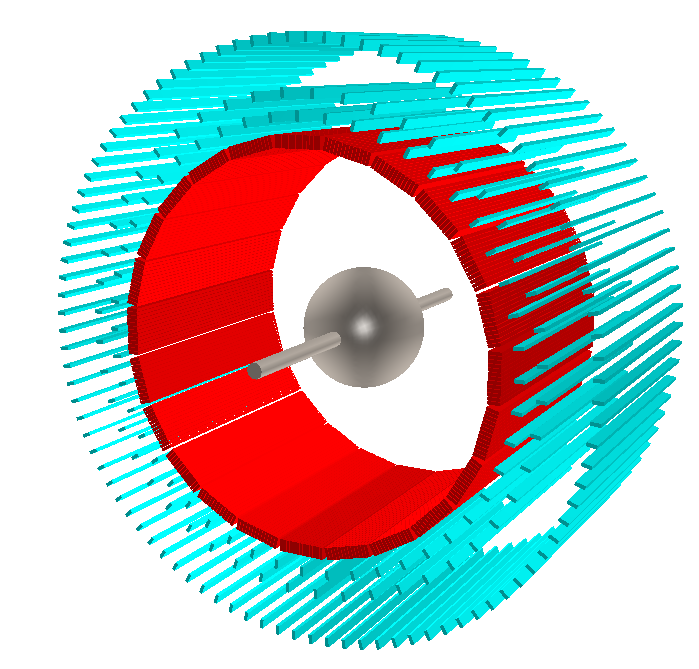}
    \caption{Left: View of the \jpet{} detector with a cylindrical vacuum chamber for positronium production and annihilation mounted in its centre.
      Right: Schematic view of two future extensions of the experimental setup:
      \textit{(i)}~the current three layers of sparsely-arranged scintillator strips~(blue) will be complemented by a layer of 24 modules containing 13 densely-packed scintillator strips each~(red); \textit{(ii)}~the cyllindrical annihilation chamber will be replaced by a spherical one~(gray).}
    \label{fig:detector}
\end{figure}

The core of the detector is constituted by 192 photon detection modules
sparsely arranged in three concentric layers along the longitudinal axis of the detector
as presented in the left panel of Figure~\ref{fig:detector}.
Each module consists of an EJ-230 plastic scintillator strip of 50~cm length
and 7$\times$19~mm$^2$ cross-section,
whose both ends are optically coupled to Hamamatsu R9800 photomultiplier tubes.

Interactions of photons in the plastic scintillators are recorded through their Compton scattering
resulting in an energy deposition depending on the scattering angle
and emission of scintillation light recorded by the two photomultipliers. 
Time of interaction and its position along the strip are determined using
time difference between light recording at the two ends of the strip.
Lack of registration of the full energy peak in \jpet{} detection modules
is compensated by an excellent interaction time resolution at the level of 100~ps~\cite{Moskal:2014sra}
resulting from fast front-end electronics~\cite{Palka:2017wms} and short decay times of plastic scintillators.
Additionally, the latter allows for pileup-free measurements with high positron source activities
of 10~MBq and more.

Electric signals from the photomultipliers are sampled in the voltage domain
at four configurable voltage thresholds by scalable front-end electronics
developed for \jpet{} coupled with a data acquisition (DAQ) system based
on FPGA chips and the TRB3 platform~\cite{Palka:2013kwa, Korcyl:2016pmt, Palka:2017wms}.
The DAQ of \jpet{} is reconfigurable and opens the possibilities of
real-time data reconstruction  directly on the FPGA systems~\cite{korcyl_ieee}.

Energy deposited by a $\gamma$ quantum interacting in a scintillator strip
is measured through total charge of the electric signals from the attached photomultipliers,
estimated by the signals' time-over-threshold sampled at the four predefined voltages.
Thanks to the ability of \jpet{} to record Compton scattering angles in multiple scattering events,
deposited energy corresponding unanimously to a given angle is related to
the recorded time over threshold allowing for calibration
of deposited energy measurement~\cite{Sharma:2019vrv, Sharma:2020}.

The scope of applications of the detector ranging from medical imaging development
to fundamental studies of positronium annihilations (including modes not observed to date such as o-Ps$\to 4\gamma$~\cite{moskal_potential}) requires the DAQ system to impose
a minimal bias on the spectrum of recorded events~\cite{Gajos:2018wyi}.
This is achieved with recording of data in a trigger-less mode~\cite{Korcyl_bams, Korcyl:2016pmt}
followed by filtering and reconstruction with a dedicated analysis software framework~\cite{Krzemien:2015hkb, Krzemien:2016hgt, Krzemien:2020ewi}.
While an unusual choice due to large resulting data volume,
in case of searches for small effects such as rare decays and symmetry violations
which can be easily mimicked by a non-uniform response of the detector and DAQ elements such as trigger bias imposed on the data,
only complete foregoing of the trigger allows for full control over systematic effects in the experiments.

Currently, \jpet{} is being extended with an additional layer of detection modules
organized in a dense layout placed within the current setup as displayed schematically in the right panel of Figure~\ref{fig:detector}.
These modules are intended to enhance the angular acceptance of the detector as well as to
provide an improved time resolution thanks to an entirely digital readout using matrices of silicon photomultipliers~\cite{Moskal:2016ztv}.
The impact of this extension of the detection setup on the discrete symmetry tests' sensitivity is discussed in Section~\ref{sec:perspectives}.

\section{Methods of searching for discrete symmetry violations with ortho-positronium in \jpet{}}\label{methods}
The ability to record photons in an energy range corresponding to electron-positron annihilations as well as below it makes \jpet{} a suitable device for studying decays of the lightest purely leptonic bound system, the positronium exotic atom.
Positronium, the bound state of electron and positron, may be formed as a singlet or triplet ground state,
referred to as para-positronium~(p-Ps) and ortho-positronium~(o-Ps) respectively. Being an antisymmetric eigenstate of charge conjugation, the latter may only annihilate into an odd number of photons due to the conservation of the C symmetry, tested to the level of~$10^{-6}$ for positronium~\cite{PhysRevA.54.1947,PhysRevA.66.052505}.
In practice, ortho-positronium predominantly annihilates into a three-photon final state with the next allowed final state~(5$\gamma$) suppressed by a factor of $\alpha^2$.

While the positronium physics appears to be well described by electromagnetic interactions
where CP violation is not expected, any observation of CP noninvariance in this system
would be an indication of new physics. Motivation for such searches is further encouraged
by the recent neutrino oscillation measurements hinting at leptonic CP violation
at 3$\sigma$ level~\cite{PhysRevLett.121.171802, PhysRevD.98.032012, Abe:2019vii},
for which no confirmation was provided by charged lepton systems to date.
As pointed out by Bernreuther and Nachtmann~\cite{Bernreuther:1988tt},
the three-photon annihilations of the triplet state of positronium
may provide insight into CP and even CPT-violating effects through certain
angular correlations between o-Ps spin and momenta of annihilation photons.

\begin{table}[H]
  \caption{Angular correlation operators constructed with observables of ortho-positronium annihilations into three photons: positronium spin $\vec{S}$ and momenta of the annihilation photons ordered by their magnitude: $|\vec{k}_1| > |\vec{k}_2| > |\vec{k}_3|$. Each of these operators is either even~(+) or odd~(--) with respect to the basic symmetry transformations and their combinations as marked in the table.
    \label{tab:operators}}
\centering
\begin{tabular}{ccccccc}
\toprule
  \textbf{no.} & \textbf{operator} & \textbf{C} & \textbf{P} & \textbf{T} & \textbf{CP} & \textbf{CPT} \\
\midrule
  1 & ${\vec{S} \cdot \vec{k_1}}$ & + & -- & + & -- & --\\      
  2 & ${\vec{S} \cdot (\vec{k_1}\times\vec{k_2})}$ & + & + & -- & + & --\\
  3 & $({\vec{S} \cdot \vec{k_1})(\vec{S} \cdot (\vec{k_1}\times\vec{k_2})})$ & + & -- & -- & -- & +\\
  \bottomrule
\end{tabular}
\end{table}

Table~\ref{tab:operators} presents three angular correlations measurable in the ortho-positronium three-photon annihilations. The correlations are represented as operators whose properties under
the C, P and T transformations and their combinations follow from
the respective behaviour of positronium spin ($\vec{S}$)
and momentum vectors of the final state photons
~($\vec{k}_i$ for $i=1,2,3$ where the photons are labeled according to
descending energy, i.e. $|\vec{k}_1| > |\vec{k}_2| > |\vec{k}_3|$ )
under these operations.
In case of operators which are antisymmetric under a given transformation (marked with ``-'' in the table), expectation value of the operator must vanish if the respective transformation constitutes a good symmetry.
Consequently, observation of a non-zero expectation value of such operator would be an indication
of violation of a given discrete symmetry~\cite{Bernreuther:1988tt, Gajos:2018wyi}.
The notion of testing discrete symmetries in the annihilations of ortho-positronium is therefore based on experimental determination of the expectation values of the angular correlation operators listed in Table~\ref{tab:operators}.
Notably, only one experiment conducted to date attempted to probe a continuous distribution of such expectation values~\cite{cpt_positronium} whereas all previous measurements were constrained to determination of an up-down asymmetry of the operators, a special case with significantly limited sensitivity~\cite{Skalsey:1991vt, Arbic:1988pv, cp_positronium}.

The \jpet{} experiment aims at precise measurements of two out of the three
presented angular correlations,
probing their full geometrically-allowed domains for the first time.

\subsection{Estimation of positronium spin}\label{sec:spin}
Essential component of the angular correlations in positronium decays
considered in this work is the knowledge of the positronium spin quantization axis.
Former measurements either used a polarized positronium beam~\cite{Arbic:1988pv},
external magnetic field~\cite{Skalsey:1991vt, cp_positronium}
or relied on the intrinsic linear polarization of positrons
emitted in $\beta^+$ decay~\cite{cpt_positronium}.
The two former approaches exclusively allow for producing
a degree of tensor polarization in the positronium sample, inevitable for
conducting a test of the CP symmetry with operator no. 3 from Table~\ref{tab:operators}.
However, setups required to convey the beam to the annihilaition recording device
and magnets providing sufficient $\vec{B}$ field effectively prevent recording
of the annihilation photons with a large angular acceptance.

Therefore, \jpet{} builds on the o-Ps polarization control scheme proposed in
the best measurement of the ${\vec{S} \cdot (\vec{k_1}\times\vec{k_2})}$ operator to date~\cite{cpt_positronium},
in which logitudinally-polarized positrons from a point-like $\beta^+$ source
of \textsuperscript{68}Ge or \textsuperscript{22}Na are allowed to form positronia
only in a limited volume which defines a range of allowed $e^+$ spin quantization axes.
As the positron polarization statistically translates to the formed ortho-positronium
in 2/3 of cases, this allows for obtaining an estimate of the o-Ps spin direction
with a finite uncertainty determined by the $\beta^+$ emission average energy
and the applied geometry of positronium formation medium.
In the original implementation of this idea, this uncertainty accounted
for a reduction of statistical polarization by 0.686, resulting in P$\approx$ 0.4
in the whole experiment~\cite{cpt_positronium}. On the other hand, it evaded the need
for a acceptance-limiting hardware setup which allowed for the first measurement
of a true distribution of an angular correlation in o-Ps annihilation, although
limited in resolution by the coarse detector granularity.

In the measurements with \jpet{} proposed in this work, we extend the idea
of estimating ortho-positronium spin without externally-induced polarization.
While it limits the accessible symmetry violating operators to positions 1. and 2.
from Table~\ref{tab:operators} as measurement of the correlation $({\vec{S} \cdot \vec{k_1})(\vec{S} \cdot (\vec{k_1}\times\vec{k_2})})$ is only possible in case of a tensor-polarized
positronium sample~\cite{MOHAMMED_2017}, this allows \jpet{} to observe an unprecedented
spectrum of angular configurations of o-Ps decays and thus the full spectra of correlation
operators 1. and 2.

To this end, we use positrons emitted from a point-like $\beta^+$ source which are characterized
by linear polarization along their direction of emission to a degree of $P_{e^+}=\upsilon/c$ where $\upsilon$ is the positron velocity and $c$ is the speed of light.
The positrons are allowed to thermalize in a layer of porous medium
enhancing positronium formation, which is spatially separated from the $\beta^+$ source
by the volume of vacuum chamber ensuring free propagation of the positrons.

In contrast to the previous measurement~\cite{cpt_positronium}, we do not assume the
positronium production region to be point-like but use the information on the locations
and times of the three photons' interactions in the detector to reconstruct
the o-Ps$\to 3\gamma$ annihilation point with a trilateration-based approach~\cite{gajos_gps}.
In consequence, we can estimate the direction of $e^+$ spin separately for each event,
thus reducing the related decrease in statistical o-Ps polarization from 0.686
to about 0.98 determined by the
spatial resolution of the o-Ps$\to 3\gamma$ reconstruction. 

In the currently performed measurements, \jpet{} implements the aforementioned spin estimation
scheme with a cylindrical positronium production chamber mounted axially in the center
and extending for the whole length of the detector.
A~10~MBq $\beta^+$ source of \textsuperscript{22}Na is installed in the center of the chamber,
while its walls are coated with 3~mm of R60G porous silica, allowing practically all positrons
reaching the chamber walls to thermalize and interact within this layer~\cite{gajos_slopos}.
The chamber walls are made of 3~mm polycarbonate so as to minimize absorption and scattering
of annihilation photons.
The chamber mounted inside the \jpet{} detector is presented in the left panel of
Figure~\ref{fig:detector}.
The right panel of the figure illustrates a future enhancement of the chamber geometry,
i.e.\ replacement of the cylinder with a spherical vacuum chamber (with R=10~cm)
which allows for a more efficient utilization of positrons from the $\beta^+$ source for positronium formation,
increases o-Ps$\to 3\gamma$ registration efficiency for extreme values of certain correlations
and reduces spurious asymmetries
as demonstrated in the next Sections. 

\subsection{The correlation between o-Ps spin and annihilation plane}\label{sec:operator2}
The \nth{2} operator from Table~\ref{tab:operators} is sensitive to potential violations
of CPT invariance and has been previously studied
in two experiments with the most precise result of the CPT-violation parameter
$C_{CPT}$ of $(2.6\pm3.1)\times 10^{-3}$~\cite{Arbic:1988pv,cpt_positronium}.
In fact, a similarly-defined triple correlation has been studied in search for T violation
in decays using $Z^0$ spin and momenta of the most energetic produced jets~\cite{PhysRevLett.75.4173}

As can be seen from Table~\ref{tab:operators}, the ${\vec{S} \cdot \vec{k_1}}$ correlation is also
odd under the CPT transformation. The choice of the ostensibly more complicated operator
in the previous measurements was motivated by the fact that ${\vec{S} \cdot (\vec{k_1}\times\vec{k_2})}$
contains a simple correlation between the o-Ps spin and positronium annihilation plane spanned by
the momentum vectors of the emitted photons. The definition using two most energetic photons' momenta
is merely an experimentally-useful convention and does not introduce a
significant correlation between detection efficiency and photon energy as is the case for
the ${\vec{S} \cdot \vec{k_1}}$ operator as argued later on in this work.
In order to avoid any dependence on selected photon energies, it is convenient to normalize
this operator to the magnitude of the cross product, leading to the following definition:
\begin{equation}
  \label{eq:operator}
  \mathcal{O}_{CPT} = \hat{S}\cdot(\vec{k}_1\times\vec{k}_2) / |\vec{k}_1\times\vec{k}_2 |,
\end{equation}
which expresses the pure angular correlation between o-Ps spin and its decay plane.

The best measurement to date was simultaneously the first measurement going beyond the up-down
asymmetry of the operator and determining its continuous distribution.
However, due to the geometry of the Gammasphere detector used therein and the positronium
production setup, the measurement was only sensitive to the values of this operator
in the range of about $(-0.4,0.4)$ out of the allowed region of [-1,1].

Due to its high granularity of the detection modules in the transverse plane and
continuous interaction position measurement in the longitudinal coordinate,
in conjunction with the spin estimation scheme which does not impose a distinguished
positronium spin quantization axis with respect to the detector,
the \jpet{} setup is able to record a substantially broader range of kinematic
configurations of o-Ps$\to 3\gamma$ events.

To demonstrate the sensitivity of \jpet{} to the distribution of the $\mathcal{O}_{CPT}$ operator,
a toy Monte Carlo simulation of ortho-positronium annihilations in the experimental setup described above
was prepared, featuring
allowed angular and energy distributions of photons from o-Ps$\to 3\gamma$ annihilations
expected from QED~\cite{lifshitz}, 
the geometry of the positronium production setup
as well as geometric arrangement of the detection modules.
Compton interactions of the annihilation photons were simulated according to the Klein-Nishina
formula and a photon registration threshold was set on the simulated deposition of energy
by the scattered photon in a scintillator strip.

The distribution of the $\mathcal{O}_{CPT}$ operator, i.e. cosine of the angle between the normal
to the annihilation plane and positronium spin direction was simulated either to be uniform
as expected in absence of CPT-violating effects~\cite{Bernreuther:1988tt} or with an assumed
level of asymmetry quantified by a $C_{CPT}$ coefficient. Following the approach used in
Ref.~\cite{cpt_positronium}, the simplest asymmetric form of the distribution as a function of
$\cos\theta$ was used where the total probability distribution contains a term linear in
$\cos\theta$ whose contribution given by $C_{CPT} \in [0;1]$.

\begin{figure}[hbt]
  \centering
  \includegraphics[width=7.5cm]{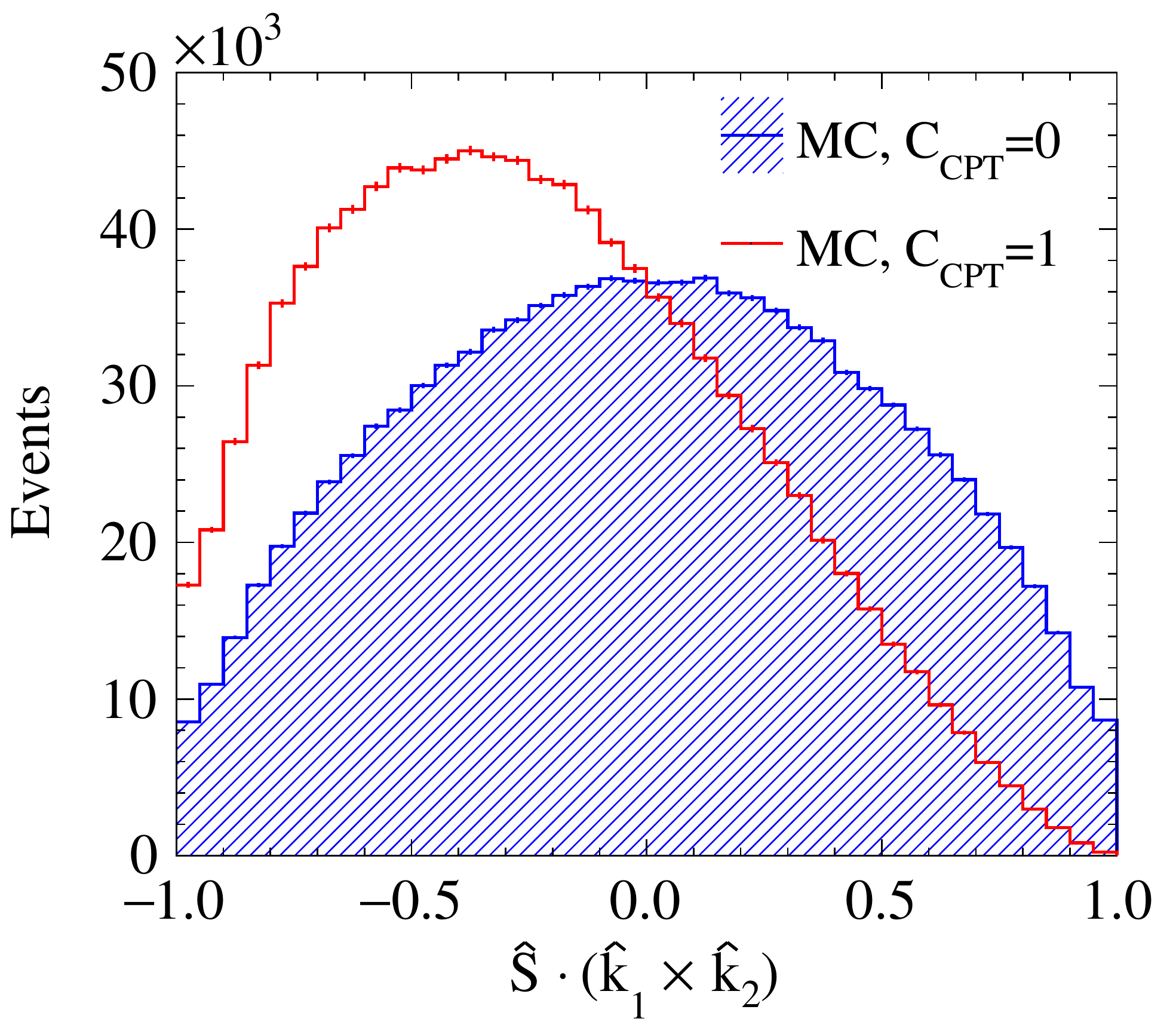}
  \caption{Distributions of the $\mathcal{O}_{CPT}$ operator resulting from toy MC simulations
    of $10^{13}$ positron interactions in \jpet{} with the cylindrical positronium production chamber
    in case of no CPT violation assumed in the simulation (hatched blue histogram)
    and extreme violation (hollow red histogram). 
  }
    \label{fig:operator}
\end{figure}

A simulation of $10^{13}$ positrons emitted from a $\beta^{+}$ source without CPT-violating effects
in ortho-positronium annihilations results in a distribution
of the $\mathcal{O}_{CPT}$ operator presented in the hatched blue histogram presented in Fig.~\ref{fig:operator}.
Red histogram in the same figure corresponds to the distribution obtained
if maximal violation of the symmetry ($C_{CPT}=1$) is assumed. While the existing  measurements
exclude values of $C_{CPT}$ beyond the $10^{-3}$ level, the exaggeration in Fig.~\ref{fig:operator}
was used to visualize the effects detectable through determination of the distribution of this
angular correlation.
It is visible that while detection efficiency peaks for values corresponding to the decay plane
normal being close to perpendicular to the spin quantization axis,
it does not drop to zero close to the extreme values of the correlation, in contrast to the previous
measurement~\cite{cpt_positronium}.

The factors determining the detection efficiency of o-Ps$\to 3\gamma$ events in \jpet{} comprise
\textit{(i)} probability of interaction of an annihilation photon in a plastic scintillator strip which
on average amounts to about 20\%,
\textit{(ii)} geometrical acceptance resulting from the sparse arrangement of detection modules and their length;
modules cover about 0.21 of the solid angle around the center of the detector,
\textit{(iii)} the energy deposition threshold above which a Compton-scattered photon
is registered by a detection module.
Furthermore, the total efficiency of observing ortho-positronium annihilations as a function of
the $\beta^+$ source activity also involves (\textit{iv}) the fraction of positrons forming o-Ps in the region
of the annihilation chamber where the three emitted photons can be recorded simultaneously.

\begin{figure}[hbt]
  \centering
  \includegraphics[width=7.5cm]{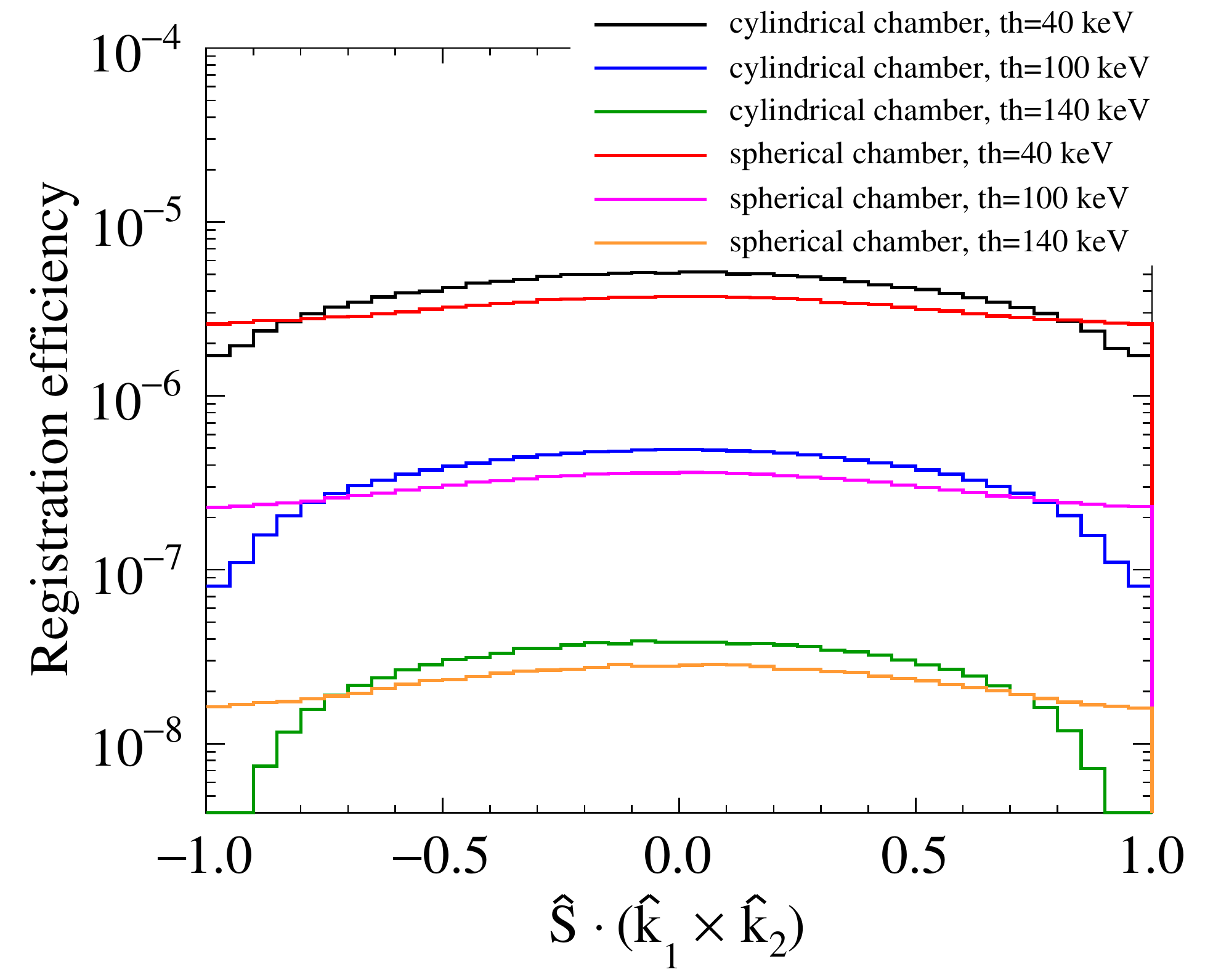}
  \includegraphics[width=7.5cm]{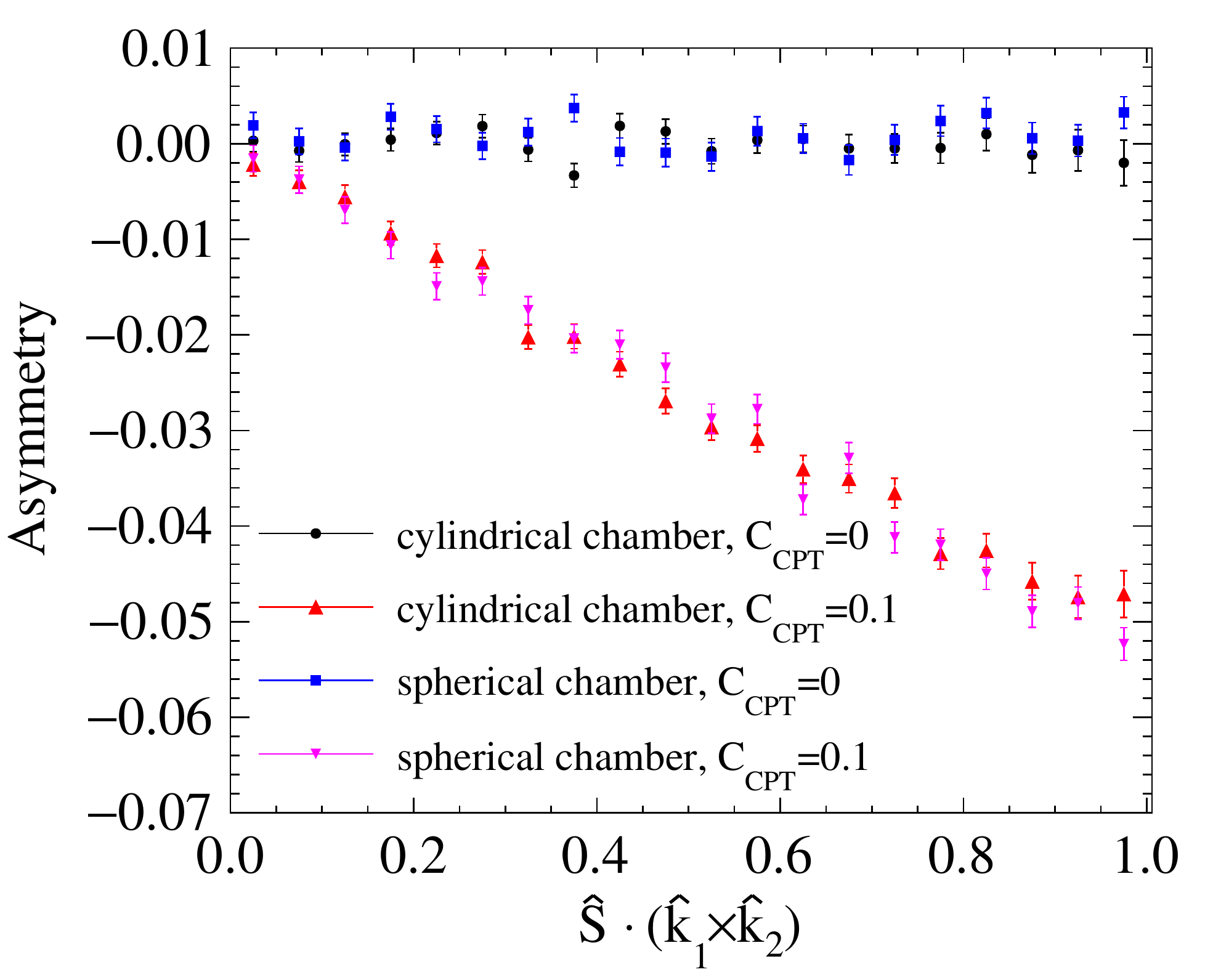}
  \caption{Left: Total efficiency of registration o-Ps$\to 3\gamma$ events in \jpet{} as a function
    of the ${\vec{S} \cdot (\vec{k_1}\times\vec{k_2})}$ angular correlation
    obtained in a MC simulation.
    The curves present efficiencies in case of two different geometries of the positronium production
    chamber: cylindrical and spherical as well as for three values of energy deposition threshold
    for single $\gamma$ detection.
    Right: Asymmetry of the ${\vec{S} \cdot (\vec{k_1}\times\vec{k_2})}$ distribution for the two
    chamber geometries in cases of no CPT violation
    and an exaggerated violation at the 10\% level assumed in the simulations.
  }
    \label{fig:operator2}
\end{figure}

Fig.~\ref{fig:operator2}~(left) presents the total $o-Ps\to 3\gamma$ registration efficiency
as a function of the angular correlation defined in Eq.~\ref{eq:operator} obtained in toy MC
simulation of $10^{13}$ positrons from a \textsuperscript{22}Na source
in the setup described in Sec.~\ref{sec:spin}. The efficiency is presented for two geometries
of the annihilation chamber: cylindrical~(presently used) and spherical~(in preparation).
In each case, three values of the energy deposition threshold for a single annihilation
photon were considered: 40~keV, 100~keV and 140~keV.
Results of the simulation show that lowering this photon registration threshold is vital
for the total efficiency and each increase of the threshold by about 50~keV results in a reduction 
of the $3\gamma$ registration efficiency by an order of magnitude.

Presently, the detection threshold of the \jpet{} photomultiplier tubes and signal sampling electronics
achievable without entering the noise level 
is estimated to be about 80~keV.

The MC-based evaluation of detection efficiency also demonstrates the enhancement expected
with the spherically-shaped positronium production vacuum chamber instead of the currently
used cylindrical one.
The drop of efficiency close to the extreme values of $\mathcal{O}_{CPT}$ visible in Fig.~\ref{fig:operator}
will be substantially reduced with the new source geometry, resulting in an efficiency more flat across
the whole operator spectrum.

A subtle asymmetry in the distribution of a given angular correlation X  may be detected by evaluation of the following figure,
accounting for asymmetry between event counts $N$ in subsequent intervals of for positive and negative values
of a given angular correlation operator $\mathcal{O}_{X}$:

\begin{equation}
  \label{eq:asym}
  A(|\mathcal{O}_X|) = \frac{N(|\mathcal{O}_{X}|) - N(-|\mathcal{O}_{X}|)}{N(|\mathcal{O}_{X}|) + N(-|\mathcal{O}_{X}|)}.
\end{equation}

Subsequently, a comparison of the $A(|\mathcal{O}_{CPT}|)$ distribution with one obtained in case of the MC-simulated distribution assuming maximal violation ($C_{CPT}$=1) would allow for extraction of the CPT symmetry
violation coefficient in a similar manner as done e.g.\ in Ref.~\cite{cpt_positronium}.
For such a procedure, good understanding of the detector efficiency as a function of the value of the measured
operator is crucial in order to avoid artificial asymmetries arising from efficiency nonuniformities
due to the setup geometry.
In \jpet{}, thanks to the large granularity of detection modules and
continuous measurement of interaction positions along them, the expected shape of such efficiency
is smooth as demonstrated in the left panel of Fig.~\ref{fig:operator2}.
While this is already an improvement with respect to the previous measurement of the $\mathcal{O}_{CPT}$ operator
where coarse granularity of the detectors constituting the Gammaspere array caused strong periodic fluctuations
of efficiency~\cite{cpt_positronium}, 
the impact of detector geometry on the measured asymmetry requires a careful treatment nonetheless.

Figure~\ref{fig:operator2}~(right) shows examples of asymmetries of the $\mathcal{O}_{CPT}$ operator
defined as in Eq.~\ref{eq:asym} using the two positronium production chamber geometries considered in this work,
for the cases of no asymmetry assumed in the MC simulations
and of CPT violation at a level of 10\%~($C_{CPT}=0.1$), exaggerated for better visibility.
These results were obtained with a simulation of $10^{13}$ positrons from a \textsuperscript{22}Na source.
It is visible that the ${\vec{S} \cdot (\vec{k_1}\times\vec{k_2})}$ angular correlation is not sensitive
to the geometry of the positronium annihilation region.
Not only are the asymmetries detected using the cylindrical and spherical chambers in good agreement,
but also the $A(\mathcal{O}_{CPT})$ distribution obtained in absence of simulated CPT violation
does not reveal signs of a false asymmetry in any of the cases.

These simulations confirm the robustness of the ${\vec{S} \cdot (\vec{k_1}\times\vec{k_2})}$
angular correlation as an observable of discrete symmetry tests.
While potentially sensitive to genuine effects of CPT violation, its definition allows to cancel out
many geometrical effects related to the measurement setup. For this reason, this correlation has been
favoured over the ostensibly simpler operator ${\vec{S} \cdot \vec{k_1}}$ in the past measurements,
even though every of the previous experiments focusing on ${\vec{S} \cdot (\vec{k_1}\times\vec{k_2})}$
was  in principle capable of studying the ${\vec{S} \cdot \vec{k_1}}$ correlation as well.
Later on, we discuss the experimental differences between these two correlations.

\subsection{The correlation between o-Ps spin and most energetic photon}\label{sec:operator1}

As discussed in the previous Section, out of the two angular correlations sensitive to discrete symmetries'
violation in absence of ortho-positronium tensor polarization (operators 1. and 2. in Table~\ref{tab:operators}),
the operator ${\vec{S} \cdot (\vec{k_1}\times\vec{k_2})}$ has already been studied in several experiments.
On the contrary, the \nth{1} operator which is a simple projection of the most energetic photon momentum
onto the direction of spin of the decaying ortho-positronium atom,
has never been measured to date despite its sensitivity to both CP and CPT-violating effects.

The reason lies in its simpler construction which makes its distribution prone to spurious effects
and thus experimentally more challenging.
While its usage as an observable of a CP and CPT test requires strict control of the impact
of the measurement setup geometry on the observed asymmetry, here we argue that
smooth efficiency curves offered by the \jpet{} detector in conjunction with detailed MC simulations
may allow for the first measurement of the ${\vec{S} \cdot \vec{k_1}}$ operator.

\begin{figure}[hbt]
  \centering
  \includegraphics[width=7.5cm]{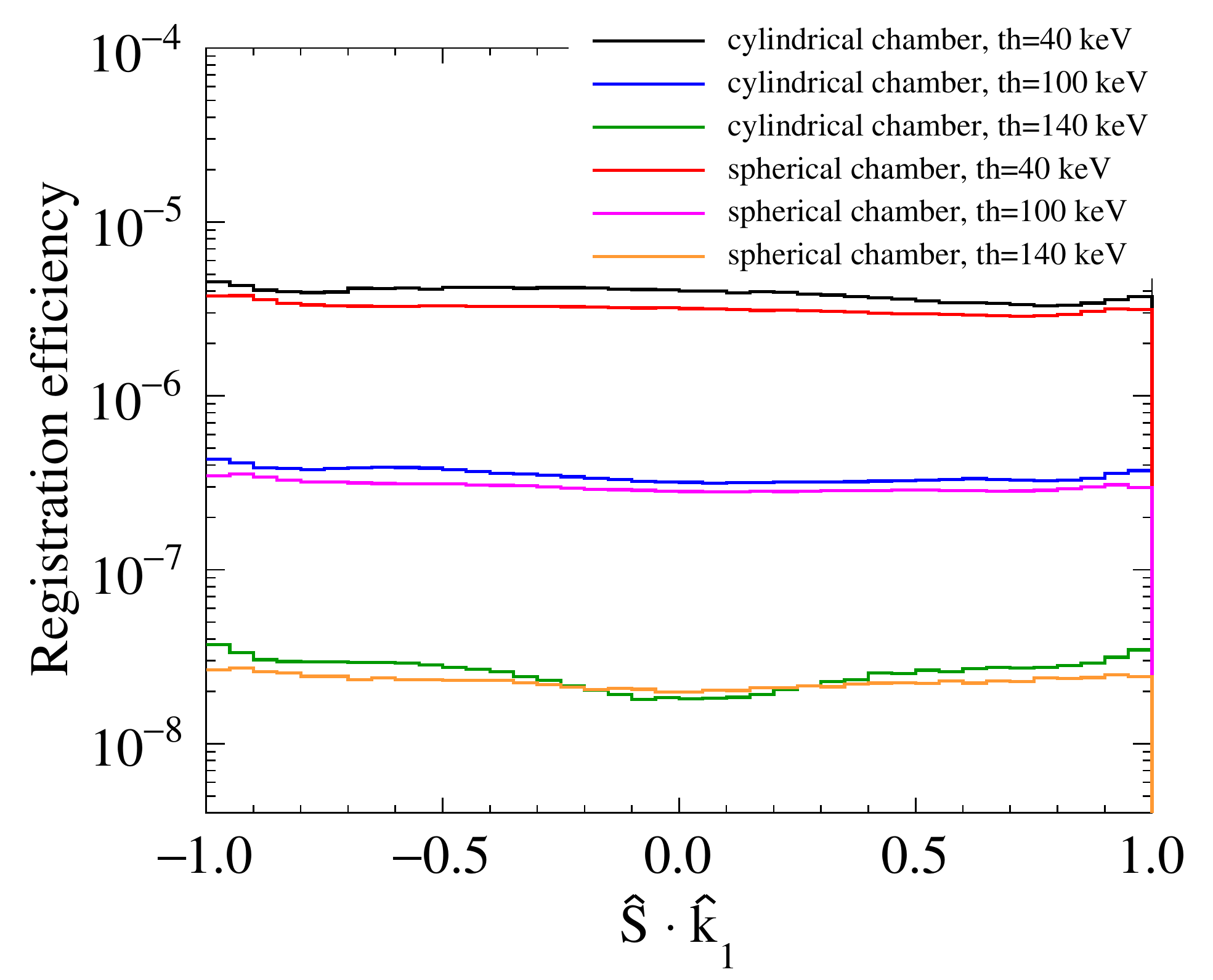}
  \includegraphics[width=7.5cm]{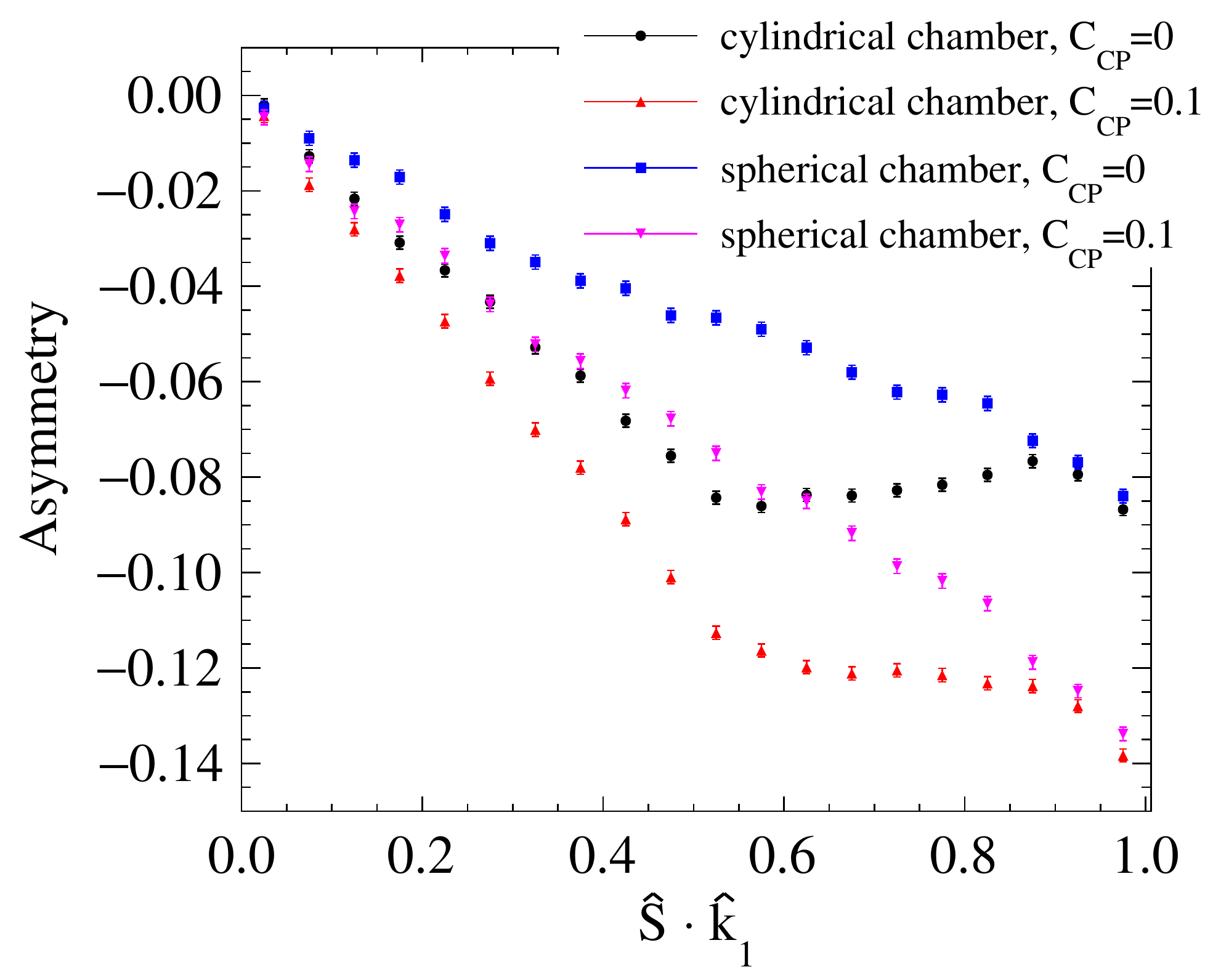}
    \caption{Left: Total efficiency of registration o-Ps$\to 3\gamma$ events in \jpet{} as a function
    of the $\vec{S} \cdot \vec{k_1}$ angular correlation
    obtained in a MC simulation.
    The curves present efficiencies in case of two different geometries of the positronium production
    chamber: cylindrical and spherical as well as for three values of energy deposition threshold
    for single $\gamma$ detection.
    Right: Asymmetry of the $\vec{S} \cdot \vec{k_1}$ distribution for the two
    chamber geometries in cases of no CPT violation
    and an exaggerated violation at the 10\% level assumed in the simulations.}
    \label{fig:operator1}
\end{figure}

Similarly as in Eq.~\ref{eq:operator}, it is convenient to introduce normalization of the
photon momentum into the definition of the angular correlation operator:

\begin{equation}
  \label{eq:1}
  \mathcal{O}'_{CPT} = \vec{S} \cdot \vec{k_1} / |\vec{k_1}|.
\end{equation}

Figure~\ref{fig:operator1}~(left) presents the efficiency of \jpet{} to o-Ps$\to 3\gamma$ events
with a given value of $\mathcal{O}'_{CPT}$ evaluated with the toy MC simulation in a similar manner
as described in Section~\ref{sec:operator2}.
A comparison with Fig.~\ref{fig:operator2}~(left) immediately reveals the challenge posed by usage
of this operator. In this case, the efficiency curves contain a modulation which is not symmetric
as a function of $\mathcal{O}'_{CPT}$.
Moreover, this effect is magnified with increasing value of the energy deposition threshold for
$\gamma$~detection.
This energy dependence originates from the choice of the most energetic photon
which introduces a correlation with the kinematical configuration of a given o-Ps$\to 3\gamma$ decay.
This phenomenon is absent in case of ${\vec{S} \cdot (\vec{k_1}\times\vec{k_2})}$ because
the geometrical entity used therein (orientation of the decay plane) is agnostic of the kinematics
of a particular annihilation event and thus also of the energy-based choice of photons.

Successful use of the $\mathcal{O}'_{CPT}$ operator as a probe of CP and CPT violation therefore
requires two factors:
\textit{(i)} maintaining the energy deposition threshold as low as possible,
\textit{(ii)} reducing the spurious asymmetries originating from asymmetric and energy-dependent
efficiency to a low and well-understood level.

The latter can be achieved by manipulation of the geometry of positronium production medium.
As displayed in the right panel of Fig.~\ref{fig:operator1},
although with both simulated setups the $\vec{S}\cdot\hat{k_1}$
a significant asymmetry appears even in case of no CPT violation assumed in the simulations 
(where possible violation is introduced in a similar manner as described in Section~\ref{sec:operator2}),
usage of the spherical vacuum chamber results in a simpler dependence of the false asymmetry
on the value of $\mathcal{O}'_{CPT}$ which is easier to parameterize.
Additionally, two independent measurements with different chambers would allow for
discrimination between the setup-specific false asymmetry and a possible genuine effect
as well as for extraction of the latter.
It is important to stress that while the results presented in this work are based on a toy MC simulation,
the actual experiments will be augmented with simulations of the full setup based on the Geant4 package,
which are currently being commissioned.

\section{Perspectives for \jpet{} sensitivity to the CPT violation effects}\label{sec:perspectives}
The \jpet{} setup featuring the cylindrical annihilation chamber is already in operation.
If a conservative photon detection threshold of 100~keV is assumed, it
can be estimated from the efficiency curve presented in the left panel of Fig.~\ref{fig:operator2}
that with a 10 MBq positron source \jpet{} can record about $8.5\times 10^{4}$ o-Ps$\to 3\gamma$
annihilations per day of measurement.
The achievable sensitivity for the CPT violation parameter $C_{CPT}$ must include
the analyzing power of the employed setup which is dominated by statistical polarization
corresponding to the estimated o-Ps spin and with a \textsuperscript{22}Na amount to about 0.4.
Taking this factor in to account, the statistical sensitivity at the unprecedented level of $10^{-4}$
can be achieved with about three months of measurement.

Although \jpet{} is well suited for extended periods of continuous measurement,
further improvements of the efficiency of both positronium production
and $3\gamma$ events detection are necessary in order to allow for reaching the sensitivity
of CPT symmetry tests discussed in this work beyond the level of $10^{-4}$.

The two upgrades already being commissioned comprise the spherical vacuum chamber for positronium
production and spin estimation and a new layer of detection modules with a fully digital readout.
The spherical chamber, in addition to the advantages discussed in Sections~\ref{sec:operator2} and~\ref{sec:operator1}, is expected to increase the fraction of positrons from the $\beta^+$ source mounted at its centre by a factor of about 1.5 with respect to the cylindrical geometry.

This is because the most sensitive region of the J-PET detector spans in the central region along its Z axis,
corresponding to $|z|<$8~cm~\cite{Gajos:2018wyi}. Outside of this volume,
registration probability for 3$\gamma$ event drops rapidly,
therefore in case of the cylindrical chamber only positrons emitted from the
$\beta^+$ source into a solid angle of about 2.2~$\pi$ have a chance to form positronium
whose annihilation can be recorded. Therefore, only 55\% of emitted positrons
may produce recordable positronium.
With the spherical geometry, over 80\% of isotropically-emitted positrons will reach the porous medium
in the most sensitive region of the detector.

The second upgrade is constituted by insertion of a new system of 312 plastic scintillator strips
arranged in 24 densely-packed modules as the innermost layer of the detector as visualized in the right
panel of Fig.~\ref{fig:detector}.
The new fully digital readout system based on silicon photomultipliers is expected to improve
time resolution of $\gamma$ interaction recording, crucial for the trilaterative reconstruction
of o-Ps annihilations and thus for the event-by-event spin estimation resolution~\cite{gajos_gps}.
Moreover, presence of the additional detection layer will increase single photon registration probability
by a factor of about 3, leading to a 27x enhancement of the total o-Ps$\to 3\gamma$ recording efficiency.

The aforementioned upgrades account for an improvement of the statistics collectable in a unit time of measurement by
about 40 with respect to the present setup.
Therefore, with the future upgraded setup, \jpet{} is expected to reach the sensitivity to $C_{CPT}$
at the level of $10^{-5}$.

It is worth mentioning that the set of discrete symmetry tests possible with \jpet{} presented in this work
is not exhaustive. A second class of angular correlation operators may be defined
using the momenta of annihilation photons and their electromagnetic polarization
rather than positronium spin direction~\cite{moskal_potential}.
Notably, neither of such correlations has been measured to date due to the incapability of the
previous positronium experiments to measure photon polarization.
Due to the detection principle based on Compton interaction, \jpet{} is the first detector
able to provide a measurement of such angular correlations involving photon electromagnetic polarization~\cite{moskal_potential, Moskal:2018pus} and experiments towards this end are already ongoing~\cite{Raj:2018kre, raj_slopos}.

\section{Conclusions}\label{sec:conclusions}
The \jpet{} detector is capable of performing tests of the CP and CPT symmetry
by determination of the distributions of the angular correlations between ortho-positronium spin
and annihilation photons in the o-Ps$\to 3\gamma$ process. Preliminary MC simulations demonstrate
that the current experimental setup may reach the sensitivity of $10^{-4}$
for the CPT violation parameter in the measurement
using the ${\vec{S} \cdot (\vec{k_1}\times\vec{k_2})}$ operator as well as set the first
measurement of the ${\vec{S} \cdot \vec{k_1}}$ operator thanks to smooth response of the
detector in function of the angular correlations and good control over spurious asymmetries.
Future upgrades of the detector and the positronium formation chamber are expected to provide
an about 40-fold increase of statistics in the same measurement time, allowing the discrete symmetry tests with \jpet{}
to reach the sensitivity of $10^{-5}$.

\vspace{6pt} 

\funding{
  This research was funded by
  The Polish National Center for Research and Development through grant
  INNOTECH-K1/IN1/64/159174/NCBR/12,
  the Foundation for Polish Science through the MPD and
  TEAM POIR.04.04.00-00-4204/17 
  programmes, the National Science Centre of Poland through grants no.
  2016/21/B/ST2/01222 and
  2017/25/N/NZ1/ 00861,
  the Ministry for Science and Higher Education through grants no.
  6673/IA/SP/2016,
  7150/E-338/SPUB/2017/1,
  7150/E-338/M/2017 and
  7150/E-338/M/2018.
}


\conflictsofinterest{The authors declare no conflict of interest.} 



\appendixtitles{no} 


\reftitle{References}

\externalbibliography{yes}
\bibliography{refs}

\sampleavailability{Monte Carlo simulations underlying the presented studies as well as data of ortho-positronium annihilations recorded with \jpet{} can be made available upon request to the authors. The software used for data analysis is available at \url{http://github.com/JPETTomography/j-pet-framework.git}}.

\end{document}